\documentclass[11pt]{article}
\usepackage{amsfonts,amsmath}
\usepackage{latexsym,epic,graphicx}

\newcommand{\Abl}{\mathcal{D}_t}
\newcommand{\Sdelta}{\underline{\delta}}

\makeatletter

\@addtoreset{equation}{section}
\makeatother

\newcommand{\be}{\begin{eqnarray}}
\newcommand{\ee}{\end{eqnarray}}
\newcommand{\nn}{\nonumber}

\renewcommand{\d}{\delta}
\newcommand{\e}{\epsilon}
\newcommand{\bi}{\bar{\imath}}
\newcommand{\bj}{\bar{\jmath}}
\newcommand{\bk}{\bar{k}}
\newcommand{\bl}{\bar{l}}
\newcommand{\DE}{DE_{10}}
\newcommand{\KDE}{K(DE_{10})}
\newcommand{\DEC}{DE_{10}/K(DE_{10})}
\newcommand{\vp}{\varphi}
\newcommand{\s}{\sigma}
\newcommand{\soso}{SO(9)\times SO(9)}
\newcommand{\p}{\partial}

\newcommand{\G}{\Gamma}
\newcommand{\Ps}{\mathtt{P}}
\newcommand{\phit}{\Phi}
\newcommand{\phis}{\phi}
\newcommand{\Qs}{\mathtt{Q}}
\newcommand{\ns}{\mathtt{n}}
\newcommand{\vs}{\mathtt{v}}
\newcommand{\cV}{\mathcal{V}}
\newcommand{\cP}{\mathcal{P}}
\newcommand{\cQ}{\mathcal{Q}}
\newcommand{\cD}{\mathcal{D}}

\renewcommand{\S}{\Sigma}

\newcommand{\de}{\text{Lie}(DE_{10})}
\newcommand{\Z}{\mathbb{Z}}
\newcommand{\car}{\mathfrak{h}}
\newcommand{\npl}{\mathfrak{n}_+}
\newcommand{\nmi}{\mathfrak{n}_-}
\newcommand{\bx}{{\bf{x}}}
\newcommand{\lra}{\leftrightarrow}
\newcommand{\eps}{\varepsilon}
\newcommand{\g}{\gamma}

\newcommand{\clAmbda}{\tilde{\Lambda}}
\newcommand{\csIgma}{\tilde{\Sigma}}
\newcommand{\lAmbda}{\tilde{\Lambda}}
\newcommand{\sIgma}{\tilde{\Sigma}}

\begin{document}

{\flushright AEI-2006-064\\[10mm]}

\renewcommand{\thefootnote}{\fnsymbol{footnote}}

\begin{center}
{\LARGE \bf Pure type I supergravity and $DE_{10}$ }\\[1cm]
Christian Hillmann\footnote[3]{E-mail: \tt christian.hillmann@aei.mpg.de} and
Axel Kleinschmidt\footnote[5]{E-mail: \tt axel.kleinschmidt@aei.mpg.de}
\\[5mm]
{\sl  Max-Planck-Institut f\"ur Gravitationsphysik\\
      Albert-Einstein-Institut \\
      M\"uhlenberg 1, D-14476 Potsdam, Germany} \\[15mm]

\begin{tabular}{p{12cm}}
\hline\\\hspace{5mm}{\bf Abstract:}
We establish a dynamical equivalence between the bosonic part of pure type
I supergravity in $D=10$ and a $D=1$ non-linear $\s$-model on the Kac--Moody
coset space $DE_{10}/K(DE_{10})$ if both theories are suitably
truncated. To this end we make use of a decomposition of $DE_{10}$
under its regular $SO(9,9)$ subgroup. Our analysis also deals partly
with the fermionic fields of the supergravity theory and we define
corresponding representations of the generalized spatial Lorentz group
$\KDE$.\\\\\hline
\end{tabular}\\[18mm]
\end{center}

\renewcommand{\thefootnote}{\arabic{footnote}}
\setcounter{footnote}{0}

\begin{section}{Introduction}

Soon after the construction of the maximally supersymmetric $D=11$
gravity theory \cite{CJS78} it was realised that this theory
exhibits exceptional hidden symmetries $E_7$ and $E_8$ upon
dimensional reduction from $D=11$ to $D=4$ and $D=3$, respectively
\cite{CrJu79,MaSchw83}. Much research has been devoted to this unexpected
feature of maximal supergravity and its relevance for string theory
\cite{Ju83,deWiNi86,HuTo95,deWiNi01,IqNeVa02}. However, already
since the early days of the study of hidden symmetries it has
been clear that also theories with non-maximal supersymmetry (or, in
fact, no supersymmetry at all) can exhibit unexpected hidden
symmetries upon dimensional reduction
\cite{Eh57,Ju81a,Cr81,Ju82a,BrMaGi88}. For the 
case of all simple and split symmetry groups $G$ in $D=3$ the
question which higher-dimensional theories give rise to the hidden
symmetry $G$ upon dimensional reduction has been answered in
\cite{CrJuLuePo99} and there are only very few groups $G$ for which the
associated (oxidised) theory has supersymmetry. One example is the
group $D_8\equiv SO(8,8)$ which in $D=3$ is the hidden symmetry of the pure
type~I supergravity theory in $D=10$ \cite{Beetal82,Ch81} (which has
half-maximal supersymmetry) after dimensional reduction. 
Further dimensional reduction
to $D=1$ was conjectured \cite{Ju81a,Ju82a} to lead to an
infinite-dimensional symmetry of hyperbolic Kac--Moody type, denoted
$\DE$, which will be defined below. As pointed out
  in \cite{KlNi04a} the conjectured symmetry $\DE$ is consistent
  with the embedding of pure type~I into the maximal theory whose
  conjectured symmetry is $E_{10}$ since $\DE$ is a proper
  subgroup of $E_{10}$.

In this paper we revisit the hidden symmetries of pure type~I
supergravity motivated by recent results concerning such
infinite-dimensional symmetries.  Near a space-like singularity it was
found that the effective and dominant dynamics of the model can be
mapped to a so-called cosmological billiard system whose massless
relativistic billiard ball bounces off the walls of a ten-dimensional
(auxiliary) billiard table \cite{DaHe01}. The location of these walls
is identical to that of the bounding walls of the fundamental Weyl
chamber of $DE_{10}$ \cite{DaHe01} (see also \cite{DaHeNi03} for a
review of cosmological billiards).

In analogy with the $E_{10}/K(E_{10})$ model
developed in \cite{DaHeNi02} for the maximally supersymmetric  case,
we here study  a $D=1$ geodesic model on the infinite-dimensional coset
space $\DEC$, extending the cosmological billiard dynamics.
The dynamical behaviour of the $\s$-model will be related to that of
the pure type~I theory. $\KDE$ refers to the (formally) maximal
compact subgroup of $\DE$ which plays the role of a generalised
spatial Lorentz group. In the context of the $E_{11}$ approach to
Kac--Moody symmetries \cite{We00,We01}, the bosonic sectors of $D=10$
type~I theories (also with abelian vector fields) have been
investigated in \cite{SchnWe04} and the equations of motion were
derived from a $DE_{11}$ analysis in the pure type~I case. The
non-maximal pure $D=5$, $N=2$ supergravity has been studied from a
Kac--Moody perspective in \cite{MiMoYa06}.

Our main result is that a truncated version of the bosonic pure type~I
equations of motion is dynamically equivalent to a truncation of the
equations of the geodesic $\s$-model on $\DEC$. The supergravity
truncation roughly involves keeping only first order spatial gradients
but arbitrary time-dependence, similar to the truncation in the $E_{10}$
correspondence for the maximally supersymmetric theory
\cite{DaHeNi02,DaNi04}. Therefore we are {\em not} performing a
dimensional reduction to $D=1$. Along the way to  demonstrating
the correspondence we rewrite the relevant parts of the  equations in
a form which is manifestly $\soso$
covariant (see \cite{KlNi04a} for an analysis of the maximal theory in
an $\soso$ formalism). For the Kac--Moody side of the correspondence
the $\soso$ covariance is straight-forward to obtain by taking a
so-called level decomposition of $\DE$ with respect to its $SO(9,9)$
subgroup which, after the transition to compact subgroups,
leads to $\soso\subset\KDE$ covariance. On the supergravity side this
requires more work and intricate redefinitions of the standard
variables.\footnote{An $SO(n)\times SO(n)$
  covariant formulation of the bosonic type~I supergravity after
  strict dimensional reduction on an $n$-torus $T^n$, i.e. discarding all
  spatial gradients, was given in \cite{MaSchw93}. Our analysis goes
  beyond this since we keep spatial gradients. For completeness, we note that $SO(n,n;\Z)$ also appears as the T-duality group of closed string theories compactified on $T^n$.}

Besides the correspondence of the bosonic equations of motion we also
study the fermionic fields of supergravity and show how they fit into
consistent (albeit unfaithful) representations of the compact subgroup
$\KDE$ of $\DE$. This, together with the bosonic dictionary, allows us
to rewrite the supersymmetry variations in a form which not only has
manifest $\soso$ covariance but also beginnings of a full $\KDE$
covariance.

Our paper is structured as followed. First we define type~I
supergravity in our conventions and introduce some field redefinitions
in section~\ref{sugra}. In section~\ref{EOMcoset}, we define the $\DEC$
$\s$-model in one dimension and work out its equations of motion in an
$SO(9,9)$ level decomposition. By comparing the two sets of equations
of motion we will derive the dictionary relevant for the dynamical
correspondence. In section~\ref{ferm} we study the supersymmetric
aspects and fermionic fields of
type~I and their relation to $\KDE$ before we close with some remarks
and future prospects in section~\ref{concl}.

\end{section}

\begin{section}{Pure type~I supergravity}
\label{sugra}

\begin{subsection}{Action and supersymmetry}

The action of $D=10$, $N=1$ supergravity \cite{Beetal82} in our
conventions reads to lowest fermion order\footnote{In what follows, we will always neglect higher order
  fermion contributions.}
\be\label{AI}
S_{\text{I}} &=& \int d^{10}x \Bigg[\frac{\hat{E}}{4}
  \bigg(\hat{R}-\frac12\p_M\phit\p^M\phit
  -\frac1{12} e^{-\phit}H_{MNP}H^{MNP}\bigg)\\
&& -\frac{i\hat{E}}{2} \left(\bar{\psi}_M\Gamma^{MNP}D_N\psi_P +
  \frac{1}{2}\bar{\lambda} \Gamma^M D_M\lambda +\frac{1}{2}\bar{\psi}_N
  \Gamma^M\partial_M\phit \Gamma^N\lambda\right)\nn\\
&&+\frac{i\hat{E}}{48}e^{-\frac{1}{2}\phit}H_{QRS}\left(\bar{\psi}_M
  \Gamma^{MNQRS}\psi_N +\bar{\psi}_N \Gamma^{QRS}\Gamma^N\lambda -
  6\bar{\psi}^Q\Gamma^R\psi^S\right)\Bigg]\nn.
\ee
Here, $\hat{E}=\det(\hat{E}_M{}^A)$ is the zehnbein determinant and
the curvature
scalar $\hat{R}$ is defined in terms of the coefficients of anholonomy
$\hat{\Omega}_{MN}{}^A$ and 
the spin connection $\hat{\omega}_{M}{}^{AB}$ via\footnote{Our index conventions are:
  $A,B,\ldots=0,\ldots,9$ are {\em flat} space-time frame indices,
  $M,N,\ldots$ are {\em curved} space-time coordinate indices whereas
  lower case   $a,b,\ldots=1,\ldots,9$ are flat spatial frame indices
  and $m,n,\ldots$
  are curved spatial coordinate indices. Frame indices are raised and lowered
  with the flat Minkowski metric
  $\eta_{AB}=\text{diag}(-1,+1,\ldots,+1)$ and we have chosen the
  Newton constant conveniently. In section~\ref{redefs},
  we will introduce
  additional indices relevant for the $\soso$ structure to be
  studied.} 
\be\nn
\hat{\Omega}_{AB}{}^C &=& 2 \hat{E}_A{}^M \hat{E}_B{}^N \p_{[M}
  \hat{E}_{N]}{}^C = \hat{\Omega}_{[AB]}{}^C,\\
\nn
\hat{\omega}_{A\,BC} &=& \frac12\left(\hat{\Omega}_{AB\,C} + \hat{\Omega}_{CA\,B} -
  \hat{\Omega}_{BC\,A}\right) = \hat{\omega}_{A\,[BC]},\\
 \nn
\hat{R}_{MN}{}^{AB} &=& 2\p_{[M}\hat{\omega}_{N]}{}^{AB} +
2\hat{\omega}_{[M}{}^{AC}\hat{\omega}_{N]\,C}{}^B,\\
\nn
\hat{R}_M{}^A &=& \hat{E}_B{}^N \hat{R}_{MN}{}^{AB},\\
\label{omega}
\hat{R} &=& \hat{E}_A{}^M \hat{R}_{M}{}^{A}.
\ee
The Lorentz covariant derivative acting on the spinors $\lambda$, $\e$ and
$\psi_M$ is
\be
D_N \psi_M &=&  \p_N \psi_M +\frac14\hat{\omega}_{N\,AB}\G^{AB}\psi_M.
\ee
and the supersymmetry variations leaving the action (\ref{AI}) invariant are
\begin{subequations}\label{SUSYvar}
\be
\d_\e\phit &=& -i\bar{\epsilon}\lambda,\label{PhiSUSY}\\
\d_\e \hat{E}_M{}^A &=& i\bar{\epsilon}\left(\Gamma^A\psi_M
  +\frac{1}{12}{\Gamma^A}_M\lambda\right),\label{ESUSY}\\
\d_\e B_{MN} &=&
  -2ie^{\frac{1}{2}\phit}\bar{\e}\left(\Gamma_{[M}\psi_{N]}
  -\frac{1}{4}\Gamma_{MN}\lambda\right),\label{BSUSY}\\
\d_\e \lambda &=& -\frac{1}{2}\Gamma^M\e \partial_M \phit  -
  \frac{1}{24}e^{-\frac{1}{2}\phit}\Gamma^{QRS}\e H_{QRS},\label{LSUSY}\\
\d_\e \psi_M &=& D_M\e
  - \frac{1}{96}e^{-\frac{1}{2}\phit}\left({\Gamma_M}^{QRS}
  - 9\delta_M^Q\Gamma^{RS}\right)\e H_{QRS}. \label{PsiSUSY}
\ee
\end{subequations}
Note that the dilatino $\lambda$ and the gravitino $\psi_M$ have opposite
spinor chirality as $SO(1,9)$ Majorana--Weyl spinors. As both can be
derived from a single eleven-dimensional gravitino,\footnote{See
  e.g. \cite{Beetal82} for the detailed derivation.} we have used
$(32\times 32)$ $\Gamma$-matrices\footnote{Our $\G$-matrix conventions can
  be found in the appendix \ref{appendix}.} $\G^A$. The $\G$-matrices $\G^M$
with curved indices appearing in (\ref{AI}) and below are obtained by
conversion with the inverse zehnbein $\G^M=\G^A\hat{E}_A{}^M$. The
spinors $\psi_M$ and
$\epsilon$ are understood as projected to one chiral half and $\lambda$ to
the other one. Spinor conjugation is defined by
$\bar{\epsilon}=\epsilon^T\G^0$. The field strength of the NSNS
two-form $B_{MN}$ is defined by $H_{MNP}=3\p_{[M}B_{NP]}$.

The purely bosonic equations of motion deduced from
(\ref{AI}) are
\begin{subequations}\label{eom1}
\be\label{eeom}
\hat{K}_{AB}&:=& \hat{R}_{AB} - \frac12\hat{\p}_A \phit \hat{\p}_B \phit
- \frac14 e^{-\phit}
   \hat{H}_{ACD}\hat{H}_B{}^{CD}\nn\\
&& \quad +\frac1{48}\eta_{AB}e^{-\phit}\hat{H}_{CDE}\hat{H}^{CDE}
   \qquad\quad= 0,\\
\label{meom}
\hat{M}_{AB} &:=& \hat{D}^C(e^{-\phit} \hat{H}_{CAB})
   \qquad\qquad\qquad\,\,\quad\quad= 0,\\
\label{deom}
\hat{S} &:=&\hat{D}_A\hat{\p}^A\phit + \frac1{12}e^{-\phit}
  \hat{H}_{CDE}\hat{H}^{CDE}\qquad= 0.
\ee
\end{subequations}
The hats on the quantities denote their projection onto an orthonormal frame
by using the zehnbein $\hat{E}_M{}^A$, e.g. $\hat{\p}_A \equiv
\hat{E}_A{}^M\p_M$. The Lorentz covariant derivative
$\hat{D}_A$ is defined with respect to the spin connection
$\hat{\omega}_{A\,BC}$ such that for example $\hat{D}_A\hat{\p}^A\phit
=\hat{\p}_A\hat{\p}^A\phit +\hat{\omega}_A{}^{AC}\hat{\p}_C\phit$.

Finally, we have the Bianchi identities
\begin{subequations}
\be\label{Bianchi1}
\hat{D}_{[A}\hat{H}_{BCD]} &=& 0,\\
\label{Bianchi2}
\hat{R}_{[ABC]D} &=& 0,
\ee
\end{subequations}
which are satisfied trivially if one substitutes in the definitions in
terms of the zehnbein $\hat{E}_M{}^A$ and the potential
$B_{MN}$. Here, it is more useful to keep them as separate equations
since they will appear separately in the correspondence with the
$\DEC$ $\s$-model.

\end{subsection}

\begin{subsection}{Redefinitions and gauge choices}
\label{redefs}

In order to make the $SO(9)\times SO(9)$ structure manifest, we fix
the following
zero shift (pseudo-Gaussian) gauge for the zehnbein {\em \`a la} ADM
\be\label{zero}
\hat{E}_M{}^A = \left(\begin{array}{cc}N&0\\0&\hat{E}_m{}^a\end{array}\right)
\ee
and then scale the spatial vielbein components and the
dilaton field according to\footnote{These redefinitions differ
  from
  the ones in \cite{KlNi04a} since we have made an additional
  conformal transformation to arrive at the action (\ref{AI}). The
  advantage of this is that also the new lapse of (\ref{DefBos2})
  below is the usual densitised lapse as in \cite{DaHeNi02}.}
\be
e_m{}^a &:=& e^{\frac14\phit} \hat{E}_m{}^a,\nn\\
e^\phis &:=& (\det(\hat{E}_m{}^a))^{-1} e^{-\frac14\phit}.\label{DefBos1}
\ee
Furthermore, we define a new (densitised) lapse function by letting
\be
\ns := (\det(\hat{E}_m{}^a))^{-1} N
  = (\det(e_m{}^a))^\frac18 e^{\frac98\phis}N  \label{DefBos2}
\ee
and set
\be
\vs_a := \hat{\omega}_{bab}e^{-\frac{1}{4}\phit} = \partial_a\phis + {e_a}^m\partial_b {e_m}^{b},
\ee
where we have used the abbreviation $\p_a\equiv
{e_a}^m\p_m$. In general, unhatted quantities in flat indices have
been projected using the new spatial neunbein $e_m{}^a$ instead of
$\hat{E}_m{}^a$. Finally, we adopt a Coulomb-type gauge for the
two-form potential by setting $B_{tm}=0$.

The bosonic fields are now combined into two sets of new
variables. The first set contains only temporal derivatives and is
given by
\begin{subequations}\label{dict}
\be
\label{dictionary1}
\Ps_{i\bar{\jmath}}&:=&{e_{i}}^m{e_{\bar{\jmath}}}^n\left(\omega_{m\,nt}
    -\frac{1}{2} H_{tmn}\right),\\
\label{dictionary2}
\Qs_{ij}&:=&{e_{i}}^m{e_{j}}^n\left(\omega_{t\,mn}
    +\frac{1}{2} H_{tmn}\right),\\
\label{dictionary3}
\Qs_{\bi\bj}&:=&{e_{\bar{\imath}}}^m{e_{\bar{\jmath}}}^n\left(\omega_{t\,mn}
   -\frac{1}{2} H_{t\,mn}\right),
\ee
\end{subequations}
whereas the second set consists of combinations of spatial derivatives
of the fields defined via
\begin{subequations}\label{dict2}
\be\label{dictionary4}
\Ps_{ijk}&:=& -3!\ns e^{-2\phis}
   {e_i}^m{e_j}^n{e_k}^p\left(\frac{1}{4}\omega_{[m\,np]}
    +\frac{1}{24}H_{mnp}\right),\\
\label{dictionary5}
\Ps_{\bi jk}&:=& -2 \ns e^{-2\phis}
    {e_{\bi}}^m{e_j}^n{e_k}^p\left(\frac{1}{4}\omega_{m[np]}
           +\frac{1}{8}H_{mnp}\right),\\
\label{dictionary6}
\Ps_{i\bj\bk}&:=& +2\ns e^{-2\phis}{e_{i}}^m{e_{\bj}}^n{e_{\bk}}^p
   \left(\frac{1}{4}\omega_{m[np]}-\frac{1}{8}H_{mnp}\right),\\
\label{dictionary7}
\Ps_{\bi\bj\bk}&:=& +3!\ns e^{-2\phis}
     {e_{\bar{\imath}}}^m{e_{\bar{\jmath}}}^n{e_{\bar{k}}}^p
     \left(\frac{1}{4}\omega_{[m\,np]}
                 -\frac{1}{24}H_{mnp}\right) ,
\ee
\end{subequations}
where $\omega$ is the spin connection with respect to the rescaled
vielbein ${e_m}^a$. Its definition is analogous to (\ref{omega}), implying
e.g. $\omega_{m\,nt}={e_{(m}}^a\partial_t e_{n)a}$.
Both the indices $i,j,\ldots$ and $\bi,\bj,\ldots$ are frame indices
taking values in the spatial directions $1,\ldots,9$, where we
identify $e_{\bi}{}^m$ and $e_i{}^m$, so that for example
$\p_{\bi}=\p_i$.  However, they
will have different transformation properties under an $\soso$ group
we now introduce. To be more precise, the unbarred indices are $SO(9)$
vector indices of the first factor, whereas the barred indices are
vector indices of the second factor. The spatial $SO(9)$ Lorentz group
is the diagonal subgroup of $\soso$ (see also \cite{KlNi04a}).
The fields $\Qs_{ij}$, $\Qs_{\bi\bj}$, $\Ps_{ijk}$ and
$\Ps_{\bi\bj\bk}$ are totally antisymmetric, whereas the mixed
$\Ps_{i\bj}$, $\Ps_{\bi jk}$ and $\Ps_{i\bj\bk}$ are only
antisymmetric in indices belonging to the same $SO(9)$ factor of
$\soso$. Repeated indices on the same level are summed over with
$\d^{ij}$ or $\d^{\bi\bj}$.

We note that the total number of components in $\Ps_{ijk}$, $\Ps_{\bi
  jk}$, $\Ps_{\bi\bj k}$ and $\Ps_{\bi\bj\bk}$ is $816$ whereas the
number of independent components of the supergravity variables
  $\omega_{m\,np}$ and $H_{mnp}$ involved in the redefinition is only
$408$ so that the
redefinition is not one-to-one. Hence, there are equivalent ways of
expressing a supergravity expression in these new variables. Our
choice is such that it connects well to the $\DE$ analysis.

\end{subsection}

\begin{subsection}{Supergravity dynamics}

We now take certain combinations of the equations of motion (\ref{eom1})
after separating the time index $0$ from the spatial indices $a$. The
independent components of the equations of motion then are
\be
\hat{K}_{ab}=0,
\qquad
&
\hat{K}_{00}  = 0,
&
\qquad
\hat{K}_{a0}  =0,
\nn\\
\hat{M}_{ab}=0,
\qquad
&
\hat{M}_{a0}  =0,
&
\qquad
\hat{S}       = 0.
\ee
We combine the symmetric Einstein equation and the antisymmetric
two-form equation into a single tensor equation with no definite symmetry
\be\label{Tens}
N^2\left(\hat{K}_{ab} -\frac{1}{4}\delta_{ab}\hat{S}
+\frac{1}{2}e^{\frac{1}{2}\phit}\hat{M}_{ab}\right)=0.
\ee
In the new variables (\ref{dict}) and (\ref{dict2}), the equation
(\ref{Tens}) takes the form
\be\label{Tensor}
&&\!\!\!\!\!\!\!\!\!\!\!\!\!\!\!\!\!
   \ns D_t\left(\ns^{-1}\Ps_{i\bar{\jmath}}\right)
   -e^{2\phis}\left(\Ps_{ikl}\Ps_{\bar{\jmath}kl}
     + 2\Ps_{\bar{k}il} \Ps_{l\bar{k}\bar{\jmath}}
     + \Ps_{i\bar{k}\bar{l}}\Ps_{\bj\bk\bar{l}}\right) \nn\\
&=& \ns \p_p \left[{e_k}^{p}\Ps_{\bar{\jmath}ki}
    -{e_{\bar{k}}}^p \Ps_{i\bar{k}\bar{\jmath}}\right]
   -e^{2\phis}\left(\Ps_{ikl}\Ps_{\bar{\jmath}kl} +
    \Ps_{i\bar{k}\bar{l}}\Ps_{\bar{\jmath}\bar{k}\bar{l}}
    -2\Ps_{\bar{k}il}\Ps_{l\bar{k}\bar{\jmath}} \right)\nn\\
&& -\ns^2e^{-2\phis}\partial_{(i} \vs_{\bar{\jmath})}
   +\ns e^{-\phis}\partial_{(i}\left[\ns
     e^{-\phis}\left(\ns^{-1}\p_{\bar{\jmath})} \ns
    -\p_{\bar{\jmath})}\phis\right) \right]
\ee
with the $\soso$ covariant derivative $D_t$ acting on $\Ps_{i\bj}$ via
\be\label{sosocov}
D_t\Ps_{i\bar{\jmath}} &:=& \p_t\Ps_{i\bar{\jmath}}
   +\Qs_{ik}\Ps_{k\bar{\jmath}} + \Qs_{\bar{\jmath}\bar{k}}\Ps_{i\bar{k}}.
\ee

The dilaton equation of motion (\ref{deom}) can be combined with the
spatial trace of the Einstein equation by
\be
    N^2 \left(\frac{1}{4}\hat{S} - \delta^{ab}\hat{K}_{ab}\right)=0
\ee
 to give the first scalar equation of motion
\be\label{Scalar1}
&&\!\!\!\!\!\!\!\!\!\!\!\!\!\!\!\!\!
\ns\partial_{t}\left(\ns^{-1} \partial_t\phis\right)
 + \frac{1}{6} e^{2\phis}\left(\Ps_{ijk}\Ps_{ijk}
 + 3\Ps_{\bar{\imath}jk}\Ps_{\bar{\imath}jk}
 + 3\Ps_{i\bar{\jmath}\bar{k}} \Ps_{i\bar{\jmath}\bar{k}}
 + \Ps_{\bi\bj\bk}\Ps_{\bi\bj\bk}\right) \nn\\
&=& \ns^2e^{-2\phis}\left[
 2\partial_a \vs_a
-\frac{1}{2}\left.\Omega^{abc}\right.\Omega_{acb}
-\vs_a \vs_a\right]\nn\\
&& -\ns e^{-\phis}\partial_{a}\left(\ns e^{-\phis}\left[\ns^{-1}\partial_a
   \ns - \partial_a\phis\right]\right) 
   +\ns^2e^{-2\phis}\vs_a \left[\ns^{-1}\partial_a
   \ns - \partial_a\phis\right].\qquad
\ee
Furthermore, we have an independent second scalar equation
\be
-N^2(\hat{K}_{00}+\delta^{ab}\hat{K}_{ab})=0,
\ee
which is proportional to the Hamiltonian constraint. In the new
variables it reads
\be\label{Scalar2}
&&\!\!\!\!\!\!\!\!\!\!\!\!\!\!\!\!
 -\left(\partial_t\phis\right)^2 + \Ps_{i\bj}\Ps_{i\bj}
 +\frac{1}{6}e^{2\phis}\left(\Ps_{ijk}\Ps_{ijk}
 +3 \Ps_{\bar{\imath}jk}\Ps_{\bar{\imath}jk}
 +3 \Ps_{i\bar{\jmath}\bar{k}} \Ps_{i\bar{\jmath}\bar{k}}
 +  \Ps_{\bi\bj\bk} \Ps_{\bi\bj\bk}\right) \nn\\
&=&  \ns^2 e^{-2\phis}\left[ 2
 \partial_{a}\vs_a
- \frac{1}{2} \Omega^{abc}\Omega_{acb}
-\vs_a\vs_a\right].
\ee
Finally, we have two vector equations stemming from the Gauss
constraint on the two-form field, $\hat{M}_{a0}=0$, and the
diffeomorphism constraint, $\hat{K}_{a0}=0$. We combine them by
\be
N e^{-\frac{1}{4}\phit}(\hat{K}_{a0}
   \pm\frac{1}{2}e^{\frac{1}{2}\phit}\hat{M}_{a0})=0
\ee
and get the two vector constraint equations
\begin{subequations}\label{Vector}
\be
\label{Vector2}
0&=&\ns\partial_{m}\left[{e_{\bar{\jmath}}}^m\ns^{-1}\Ps_{i\bar{\jmath}}\right]
   +2\ns^{-1}e^{2\phis}\Ps_{\bar{\jmath}ki} \Ps_{k\bar{\jmath}}
   +\ns e^{-\phis}\partial_{i}\left(\ns^{-1}e^{\phis} \partial_t\phis\right),\\
\label{Vector1}
0&=&\ns\partial_{m}\left[{e_j}^{m}\ns^{-1}\Ps_{j\bar{\imath}}\right]
 -2\ns^{-1}e^{2\phis}\Ps_{j\bar{k}\bar{\imath}}\Ps_{j\bar{k}}
+\ns e^{-\phis}\partial_{\bar{\imath}}\left(\ns^{-1}e^{\phis}
  \partial_t\phis\right).  \quad\quad
\ee
\end{subequations}
The equations (\ref{Tensor})--(\ref{Vector}) are completely
equivalent to the set of bosonic equations of motion (\ref{eom1}) upon substitution of the definitions (\ref{dict}) and (\ref{dict2}).

We conclude this section by giving the Bianchi identities
(\ref{Bianchi1}) and (\ref{Bianchi2}) in an appropriate form. Starting
from the equations
\begin{subequations}\label{P2}
\be
\ns e^{-2\phis}D_t \left(\ns^{-1}e^{2\phis}\Ps_{ijk}\right)
     + 3\Ps_{[i|\bar{l}}      \Ps_{\bar{l}|jk]} &=&
    -\frac{3}{2}\partial_{[i} \Qs_{jk]}\label{P2ijk},\\
\ns e^{-2\phis}D_t \left(\ns^{-1}e^{2\phis}\Ps_{\bar{\imath}jk}\right)
       +\Ps_{l\bar{\imath}}\Ps_{ljk}
      +2\Ps_{[j|\bar{l}|}\Ps_{k]\bar{\imath}\bar{l}}
       &=& -\frac{1}{2}\partial_{\bar{\imath}}\Qs_{jk}
      +\partial_{[j}\Ps_{k]\bar{\imath}}\label{P2bijk},\\
\ns e^{-2\phis}D_t\left(\ns^{-1}e^{2\phis}\Ps_{i\bar{\jmath}\bar{k}}\right)
    +\Ps_{i\bar{l}}\Ps_{\bar{l}\bar{\jmath}\bar{k}}
    +2\Ps_{l[\bar{\jmath}}\Ps_{\bar{k}]il}
    &=&\frac{1}{2}\partial_{i}\Qs_{\bar{\jmath}\bar{k}}
    -\partial_{[\bar{\jmath}}\Ps_{|i|\bk]}\label{P2ibjbk},\quad\quad\quad\\
\ns e^{-2\phis} D_t\left(\ns^{-1}e^{2\phis}\Ps_{\bi\bj\bk}\right)
    +3\Ps_{l[\bar{\imath}}  \Ps_{|l|\bar{\jmath}\bar{k}]}
    &=&\frac{3}{2}\partial_{[\bar{\imath}}
            \Qs_{\bar{\jmath}\bar{k}]}\label{P2bibjbk},
\ee
\end{subequations}
one recovers the Bianchi identities by taking suitable
combinations. For the Bianchi identity $\hat{D}_{[0}\hat{H}_{abc]}=0$
one has to sum (\ref{P2ijk}) and (\ref{P2bibjbk}), whereas
$\hat{R}_{[0bcd]}=0$ corresponds to the difference between (\ref{P2ijk})
and (\ref{P2bibjbk}). The difference between (\ref{P2bijk}) and
(\ref{P2ibjbk}) gives the identity $R_{[0ab]c}=0$. The Bianchi
identities with purely spatial indices will not be discussed here but
they can also be rewritten in the new variables of (\ref{dict}) and
(\ref{dict2}).

\end{subsection}

\end{section}

\begin{section}{The geodesic $\DEC$ coset model}
\label{EOMcoset}

\begin{subsection}{Abstract derivation of the equations of motion}\label{deriv}

The abstract $D=1$ $\s$-model on any group coset $G/H$ is given in
terms of a representative $\cV(t)\in G/H$, where $t$ is the
parameter along the world-line.  The velocity along this worldline
pulled backed to the identity is the $\text{Lie}(G)$ valued expression
$\p_t\cV\,\cV^{-1}$ that can be
decomposed into generators along $\text{Lie}(H)$ and $\text{Lie}(G/H)$
as
\be\label{vel}
\p_t \cV\,\cV^{-1} = \cQ + \cP,
\ee
where $\cQ\in\text{Lie}(H)$ are the unbroken gauge connections in the
language of non-linear realisations, and $\cP\in\text{Lie}(G/H)$
correspond to the velocity components in the direction of the `broken'
generators. Using the invariant symmetric form ($\equiv$ Cartan--Killing
form)\footnote{This is the invariant trace in the adjoint
  representation in the finite-dimensional case.}
$\langle\cdot|\cdot\rangle$ on $\text{Lie}(G)$, we define a Lagrange
function that determines the dynamics of the bosonic
$D=1$ $\s$-model
\be\label{l}
L = \frac14 n^{-1} \langle\cP|\cP\rangle.
\ee
The Lagrange function is invariant under the standard non-linear
transformation $\cV(t)\to
h(t) \cV(t) g^{-1}$ for local $h(t)\in H$ and global $g\in G$. The
factor $n(t)$ ensures 
reparametrisation invariance along the world-line and, since we have
no mass term in $L$, the massless particle will move on a null
trajectory.\footnote{This is only possible if the invariant form
  $\langle\cdot|\cdot \rangle$ is {\em indefinite} as in our case.}

In order to derive the equations of motion from this Lagrange function
we consider variations of the field $\cV$ associated with a derivation
$\d$ which is assumed to commute with time derivative $\p_t$. Under
this variation we get a similar decomposition $\d\cV\,\cV^{-1}= \S+\Lambda$
for $\S\in\text{Lie}(H)$ and $\Lambda\in\text{Lie}(G/H)$. Substituting this
variation into the Lagrange function (\ref{l}) leads to the
$H$ covariant $\s$-model equations of motion
\be\label{sigmaeom}
n \p_t(n^{-1} \cP) - \left[\cQ,\cP\right] = 0
\ee
and the null constraint\footnote{We assume here that $\cV$ and $n$ are
  independent.}
\be\label{nullcons}
\langle\cP|\cP\rangle = 0.
\ee
Using the $H$ covariant derivative
\be\label{covder}
\cD = \p_t -\cQ,
\ee
where $\cQ$ acts on a $H$ representation, here the algebraic coset
$\text{Lie}(G/H)$, eq.~(\ref{sigmaeom}) can also be written as
\be
\cD(n^{-1}\cP) =0 .
\ee

For any Kac-Moody algebra \cite{Ka90}, there is a generalized
transposition map $-\omega$ mapping the Chevalley generators $e_i$,
$f_i$ and $h_i$ with $i=1,\dots,\text{rk}(\text{Lie}(G))$ to
themselves by
\be
-\omega(e_i) = f_i\quad,\quad -\omega(f_i) = e_i \quad,\quad -\omega(h_i) = h_i.
\ee
As $\omega^2=\mathbf{1}_{\text{Lie}(G)}$, we can decompose any
$\text{Lie}(G)$-valued object into eigenspaces $\cQ\in\text{Lie}(H)$
and $\cP\in \text{Lie}(G/H)$ via
\be
-\omega(\cP) = \cP  &\quad,\quad&    -\omega(\cQ) = -\cQ.
\ee
$H$ is then referred to as the maximal compact\footnote{In the case of a
  finite-dimensional Lie group $G$, this is the standard notion of
  a compact manifold.} subgroup of $G$, which we denote by $K(G)$.
Now we study this general set-up for the case of $G=\DE$ and $H=\KDE$.

\end{subsection}

\begin{subsection}{The $D_9$ level decomposition of $\de$}
\label{BWGL2}

\begin{figure}
\begin{center}
\scalebox{1}{
\begin{picture}(300,60)
\put(5,-5){$1$} \put(45,-5){$2$} \put(85,-5){$3$}
\put(125,-5){$4$} \put(165,-5){$5$} \put(205,-5){$6$}
\put(245,-5){$7$} \put(285,-5){$8$} \put(260,45){$9$}
\put(100,45){$10$} \thicklines
\multiput(10,10)(40,0){8}{\circle*{10}}
\multiput(15,10)(40,0){7}{\line(1,0){30}}
\put(250,50){\circle*{10}} \put(250,15){\line(0,1){30}}
\put(90,50){\circle{10}}\put(90,15){\line(0,1){30}}
\end{picture}}
\caption{\label{de10dynk}\sl Dynkin diagram of $\text{Lie}(DE_{10})$ with
numbering  of nodes. The solid nodes mark the $\mathfrak{so}(9,9)\equiv
\text{Lie}(D_9)$
subalgebra.}
\end{center}
\end{figure}
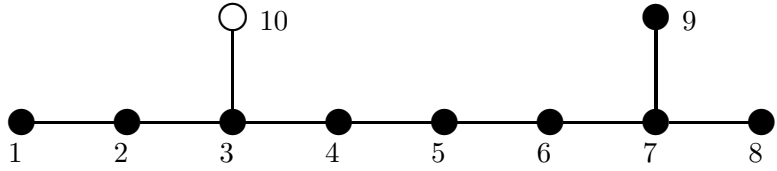

The Lie algebra $\de$ is an infinite-dimensional hyperbolic Kac--Moody
algebra \cite{Ka90} with Dynkin diagram given in
figure~\ref{de10dynk} and we consider it in split real form.
In order to analyse the dynamical equation
(\ref{sigmaeom}), we need to know the structure constants of
$\de$. However, a closed representation of $\de$ is not known.
The only known presentation of $\de$ is in terms of
simple generators $e_i$, $f_i$ and $h_i$ (for $i=1,\ldots,10$) and
defining relations among them \cite{Ka90}. These simple generators and their independent
multiple commutators form a basis of the vector spaces $\npl$, $\nmi$ and  the Cartan subalgebra $\car$ respectively. Thus, one obtains the following decomposition of $\de$:
\be
\de= \nmi \oplus \car \oplus \npl.
\ee
Next, we define the level $\ell$ of a
homogeneous element of $\npl$ with respect to
$D_9\equiv\text{Lie}(SO(9,9))$ to be the number of times $e_{10}$
appears in the corresponding multiple commutator. This definition can
be extended to the entire Kac--Moody algebra by counting $f_{10}$
negatively and setting the degree of $\car$ to zero. Thus, we
have constructed an integer grading of $\de$, given by the level
decomposition into subspaces labelled by levels $\ell\in\Z$. The level
$\ell$ piece in this decomposition is finite-dimensional and
is mapped to itself under the adjoint action of the $\ell=0$
piece. Therefore any fixed level $\ell$ is a sum of irreducible
representations of the $\ell=0$ subalgebra of $\de$ and we first study
$\ell=0$.\footnote{Further details on this
  decomposition can be found in \cite{KlNi04a}; the general technique
  of level decompositions is explained for example in
  \cite{DaHeNi02,GaOlWe02,NiFi03,KlSchnWe04}.}

Following from this definition, the subspace with level $\ell=0$
consists of all multiple generators of $e_i$, $f_i$ for $i=1,\ldots,9$
and all ten Cartan subalgebra elements. Leaving out all $e_{10}$ and
$f_{10}$ generators in commutators one arrives at a $\text{Lie}(D_9)$
subalgebra of $\de$ as also evident from figure~\ref{de10dynk}. A
certain linear combination of the ten Cartan elements $h_i$ is
orthogonal to this $\mathfrak{so}(9,9)$ and therefore the resulting
$\ell=0$ subalgebra of $\de$ is the direct sum
$\mathfrak{so}(9,9)\oplus\mathfrak{gl}(1)$.

We denote the
$SO(9,9)$ generators by $M^{IJ}=-M^{JI}$ and take their commutation
relation to be
\begin{eqnarray}
\left[M^{IJ},M^{KL}\right] =
\eta^{KI}M^{JL}-\eta^{KJ}M^{IL}-\eta^{LI}M^{JK}+\eta^{LJ}M^{IK}
\end{eqnarray}
with
\be\label{eta}
        \eta^{IJ}=\text{diag}(\mathbf{1}_9,-\mathbf{1}_9)
        &\Longleftrightarrow& \eta^{ij}=\delta^{ij} =
        -\eta^{\bar{\imath}\bar{\jmath}} \,;\,
        \eta^{i\bar{\jmath}} = \eta^{\bar{\imath}j} =0,
\ee
where we made use of the $\soso$-indices $I\equiv
(i,\bar{\imath})$.\footnote{The explicit mapping between the $D_9$ Chevalley
  operators and the $M^{IJ}$ can be found in \cite{KlNi04a}.} We use
$\eta_{IJ}$ to raise and lower $SO(9,9)$ indices in the standard
fashion. With the Cartan--Killing form
\be\label{MM}
    \left\langle M^{IJ}|M^{KL}\right\rangle =
    \eta^{KJ}\eta^{IL}-\eta^{KI}\eta^{JL},
\ee
we can split the generators into compact and non-compact ones
\be\label{J0}
J^{ij}:=M^{ij} \quad,\quad
J^{\bar{\imath}\bar{\jmath}}:=-M^{\bar{\imath}\bar{\jmath}}\quad;\quad
S^{i\bar{\jmath}}:=M^{i\bar{\jmath}},
\ee
where $J^{ij}$ and $J^{\bar{\imath}\bar{\jmath}}$ generate the two
$SO(9)$-groups of $\soso$, whereas the symmetric $S^{i\bj}$ is a coset
generator. The generator of $GL(1)$ will be denoted
$T$ and satifies
\be
    \left[T,M^{IJ}\right]&=&0\nn,\\
    \langle T|T\rangle &=& -1\label{TT}.
\ee

Restricting $\de$ to level $\ell=1$ constitutes an irreducible
representation of $\text{Lie}(D_9\times GL(1))$ which is in an
antisymmetric three-tensor representation of $SO(9,9)$, denoted by
$E^{IJK}$, and carries $GL(1)$ charge $+1$:
\begin{subequations}
\be
    \left[T,E^{IJK}\right]&=&E^{IJK},\label{TEIJK}\\
    \left[M^{IJ},E^{KLM}\right]&=& 3\eta^{I[K}E^{LM]J}
    -3\eta^{J[K}E^{LM]I}\label{MEIJK}.
\ee
\end{subequations}
The commutation relations for the elements $F_{IJK}:=-\omega(E^{IJK})$
on level $\ell=-1$ are obtained by using the generalized transposition
$-\omega$ on (\ref{TEIJK}) and (\ref{MEIJK}) to give
\begin{subequations}
\be
    \left[T,F^{IJK}\right]&=&-F^{IJK},\label{TFIJK}\\
    \left[M^{IJ},F^{KLM}\right]&=& 3\eta^{I[K}F^{LM]J}
        -3\eta^{J[K}F^{LM]I}\label{MFIJK}.
\ee
\end{subequations}
Here, indices have been raised with $\eta^{IJ}$. If we fix the
normalization by\footnote{We use $\delta^{I_1\cdots
    I_l}_{L_1\cdots L_l} :=
  \delta^{[I_1}_{L_1}\cdots\delta^{I_l]}_{L_l}$ and
  antisymmetrisations of strength one.}
\begin{eqnarray}
    \left\langle E^{IJK}|F_{LMN}\right\rangle&:=&
    12\delta^{IJK}_{LMN}\label{EF},
\end{eqnarray}
the invariance of the Cartan--Killing form yields the $\de$
commutation relation
\be
    \left[  E^{IJK} ,F_{LMN}\right]  &=& -12\delta^{IJK}_{LMN}T
    -36\delta^{[IJ}_{[LM}\eta_{N]P}M^{K]P}.\label{S2Komm}
\ee
In summary, we have the following set of generators of $\DE$ on levels
$|\ell|\le 1$ in a decomposition with respect to $D_9$:
\begin{center}
\begin{tabular}{r|c|c|c}
&$\ell=-1$&$\ell=0$&$\ell=1$\\
\hline
Generator&$F_{IJK} $& $M^{IJ}\,,\, T $&$ E^{IJK}$\\
Dimension&816 & 153+1 & 816
\end{tabular}
\end{center}
As we will not use any further generators in this paper, we will not
discuss the representation theory of the higher levels. More details
and extensive tables up to $\ell=5$ can be found in \cite{KlNi04a}.

\end{subsection}

\begin{subsection}{$\s$-model equation truncated at $D_9$ level $1$}

Given the knowledge of the generators up to level $\ell=1$ we can
parametrise the $\DEC$ coset element $\cV$ by
\be
\cV(t) = e^{\vp(t)\,T}e^{\frac12 v_{IJ}(t)\,M^{IJ}}e^{\frac1{3!}
  A_{IJK}(t)E^{IJK}} \cdots
\ee
in a Borel gauge consisting of only generators of levels
$\ell\ge 0$. Singling out the $GL(1)$-generator $T$ will introduce a
factor $e^{\ell\vp}$ for the level $\ell$ term when evaluating the
velocity (\ref{vel}), which we truncate for all $\ell\geq 2$.\footnote{
  In principle, we should add all higher level $\ell$
  $D_9$-representations $E^{(\ell)}$ with appropriate coefficients
  $P^{(\ell)}$ to the parametrization. However, in the present
  discussion we want to restrict the levels to $\ell\le 1$, which is
  consistently achieved by setting all coefficients $P^{(\ell)}=0$
  initially for all $\ell\geq 2$ \cite{DaNi04}.} Hence, we get the
parametrisation $\p_t \cV\,\cV^{-1} = \cP +\cQ$ with
\be\label{pq}
\cP &=& \p_t\vp T + P_{i\bj}S^{i\bj}
     +\frac{1}{3!}e^{\vp} P_{IJK}S^{IJK},\nn\\
\cQ &=& \frac{1}{2}Q_{ij}J^{ij}  +\frac{1}{2}Q_{\bi\bj}J^{\bi\bj}
   +\frac{1}{3!}e^{\vp} P_{IJK}J^{IJK},
\ee
where we have introduced a new notation for the coset generator
$S^{IJK}$ and the $K(DE_{10})$-generator $J^{IJK}$
\begin{subequations}
\be
    S^{IJK} &:=& \frac{1}{2}\left(E^{IJK}-\omega(E^{IJK})\right)\label{SIJK},\\
    J^{IJK} &:=&
    \frac{1}{2}\left(E^{IJK}+\omega(E^{IJK})\right)\label{JIJK}.
\ee
\end{subequations}
The occurrence of the same coefficient $P_{IJK}$ in $\cP$ and $\cQ$ in
(\ref{pq}) is due to our Borel gauge condition. Although we
could work out $P_{IJK}$ explicitly in terms of the coset
coordinate fields $v_{IJ}$ and $A_{IJK}$ we will leave it in this compact form
since this will be sufficient for the comparison with
the $\s$-model equations of motion (\ref{sigmaeom}).

In section \ref{deriv}, we mentioned that the $\s$-model equations of
motion (\ref{sigmaeom}) are only $H=K(DE_{10})$-covariant, whereas the
$G=DE_{10}$-covariance is broken. However, $SO(9,9)$ is not a subgroup
of $\KDE$, only its maximal compact subgroup $SO(9)\times SO(9)$ is.
Hence for our truncation chosen, we can only expect to get
$\soso$ covariant equations of motion. This is already obvious from
the definitions of the generators (\ref{SIJK}) and (\ref{JIJK}):
$S^{IJK}$ and $J^{IJK}$ do not transform as $SO(9,9)$ tensors, but as
$\soso$ tensors. Therefore, we also decompose the $SO(9,9)$ tensor
$P_{IJK}$ into its irreducible $\soso$ components $P_{ijk}$, $P_{\bi
  jk}$, $P_{i \bj\bk}$ and $P_{\bi \bj\bk}$ and write the $\s$-model
equations of motion (\ref{sigmaeom}) for the level $\ell=0$,
{\em i.e.} projected on the coset generators $T$ and $S^{i\bj}$,
as
\begin{subequations}\label{eoml0}
\begin{eqnarray}
     0&=& n\partial_t \left(n^{-1}\partial_t\vp \right)\nn\\
      &&\quad+\frac{1}{6}e^{2\vp}\left(P_{ijk}P_{ijk} +3P_{\bi
     jk}P_{\bi jk} +3P_{i\bj\bk}P_{i\bj\bk}
     +P_{\bi\bj\bk}P_{\bi\bj\bk}\right)\label{CosScalar1},  \\
    0&=& n D_t \left(n^{-1}P_{i\bar{\jmath}}\right)
         -e^{2\vp}\left(P_{ikl} P_{\bar{\jmath}kl} +2P_{\bk il}
     P_{l\bk\bar{\jmath}}
     +P_{i\bk\bl} P_{\bar{\jmath}\bk\bl}\right).\label{CosTensor}
\end{eqnarray}
\end{subequations}
Here, $D_t$ is the $\soso$ covariant derivative of (\ref{sosocov}) with
the connection $(Q_{ij},Q_{\bi\bj})$. The $\s$-model equations of
motion (\ref{sigmaeom}) for the level
$\ell=1$, {\em i.e.} projected on the generators $S^{ijk}$, $S^{\bi jk}$,
$S^{i\bj \bk}$ and $S^{\bi\bj\bk}$,  are
\begin{subequations}\label{eoml1}
\begin{eqnarray}
n e^{-2\vp} D_t \left(n^{-1} e^{2\vp} P_{ijk}\right)
+3P_{[i|\bar{l}}P_{\bar{l}|jk]}\label{CosP2ijk} &=& 0,\\
n e^{-2\vp} D_t \left(n^{-1} e^{2\vp} P_{\bi jk}\right)
+P_{l\bi}P_{ljk}+2P_{[j|\bl}P_{k]\bi\bl}\label{CosP2bijk} &=& 0,
\\
n e^{-2\vp} D_t \left(n^{-1} e^{2\vp} P_{i\bar{\jmath}\bar{k}}\right)
+P_{i\bar{l}}P_{\bar{l}\bar{\jmath}\bar{k}}
+2P_{l[\bar{\jmath}}P_{\bar{k}]il} &=& 0 \label{CosP2ibjbk},\\
n e^{-2\vp} D_t \left(n^{-1} e^{2\vp}
P_{\bar{\imath}\bar{\jmath}\bar{k}}\right)
+3P_{l[\bar{\imath}}P_{l|\bar{\jmath}\bar{k}]} &=& 0\label{CosP2bibjbk}.
\end{eqnarray}
\end{subequations}
As we set $P^{(\ell)}=0$ for $\ell\geq 2$ initially, the equation
(\ref{sigmaeom}), describing its time evolution, preserves this
setting. However, written in terms of the field $\cV$
parametrising the coset $\DEC$ this implies a non-trivial time
evolution of the higher level fields. We stress that the terms
extending the $\soso$ covariant derivative $D_t$ in (\ref{eoml1})
are the next terms in the full $\KDE$ covariant derivative $\Abl$
of (\ref{covder}).

We conclude this section with the null
constraint $\langle\cP|\cP\rangle =0$, cf. ~(\ref{nullcons}), in this
parametrisation
\be
0&=&  -\left(\partial_t\vp\right)^2
     +P_{i\bar{\jmath}}P_{i\bar{\jmath}}\nn\\
     &&\quad +\frac{1}{6}e^{2\vp}\left(P_{ijk}P_{ijk} +3P_{\bi
       jk}P_{\bi jk} +3P_{i\bj\bk}P_{i\bj\bk}
     +P_{\bi\bj\bk}P_{\bi\bj\bk}\right)\label{CosScalar2}.
\ee
\end{subsection}

\begin{subsection}{Comparison of the $\s$-model with supergravity}\label{Comp}

Now we turn to the comparison of the level $\ell=0$ $\s$-model
equations of motion (\ref{eoml0}) with the rewritten dynamical
supergravity equations (\ref{Tensor}) and (\ref{Scalar1}) and of the
$\ell=1$ equations (\ref{eoml1}) with the Bianchi constraints
(\ref{P2}). We will also compare the Hamiltonian constraint
(\ref{Scalar2}) with the null constraint (\ref{CosScalar2}). All these
equations contain the following objects:
\begin{center}
\begin{tabular}{c|c}
    $D=10$ pure type~I supergravity & 
  $DE_{10}/K(DE_{10})$ $\s$-model\\
    \hline
    $\Ps_{i\bj}(t,\bx)$,\, $\Qs_{ij}(t,\bx)$,\,
         $\Qs_{\bi\bj}(t,\bx)$
    & $P_{i\bj}(t)$,\, $Q_{ij}(t)$,\, $Q_{\bi\bj}(t)$\\
    $\Ps_{ijk}(t,\bx)$,\, $\Ps_{\bi jk}(t,\bx)$,\,
         $\Ps_{i\bj\bk}(t,\bx)$,\,$\Ps_{\bi\bj\bk}(t,\bx)$
    & $P_{ijk}(t)$,\, $P_{\bi jk}(t)$,\, $P_{i\bj\bk}(t)$,\,
    $P_{\bi\bj\bk}(t)$\\
    $\phis(t,\bx)$ & $\vp(t)$\\
    $\ns(t,\bx)$   &  $n(t)$
\end{tabular}
\end{center}
Here, we have explicitly re-instated the dependence on the
coordinates. Working locally in one coordinate chart $(t,\bx)$ and
keeping the spatial point $\bx$ fixed,
the time coordinate $t$ of supergravity can be identified with the
parameter along the world-line of the coset model, as already
anticipated in the table above.

By comparing the supergravity equations and the $\s$-model equations,
we see that we can match large parts of the equations by demanding
that the supergravity quantities evaluated at the fixed spatial point
$\bx$ correspond to the $\s$-model quantities. In other words, the
{\em dynamical dictionary} which maps (parts of) the supergravity equations
to the $\s$-model equations consists of letting 
\be
\Ps_{i\bj}(t,\bx)\lra P_{i\bj}(t)\quad\text{for all $t$ (and fixed
  $\bx$)}
\ee
and similarly for the other objects in the table.
To be more precise the $\s$-model equations
(\ref{eoml0})--(\ref{CosScalar2}) coincide with the left hand sides of
the equations of motion (\ref{Scalar1}), (\ref{Tensor}) and
(\ref{Scalar2}) and the Bianchi identity (\ref{P2}) of $D=10$ pure
supergravity. However, the terms on the right hand sides do not match
in this correspondence which we now discuss in more detail, together
with the vector constraint
equations (\ref{Vector2}) and (\ref{Vector1}) which do not have
corresponding $\s$-model equations.

We begin with the tensor equation (\ref{Tensor}) and the two vector
constraints (\ref{Vector2}) and (\ref{Vector1}), where we want to show
that, in some sense, we have only neglected spatial derivatives.
Our identification fixed an arbitrary spatial position $\bx$ in
a coordinate chart and considered the evolution of the fields $\Ps$ in
time only. However, direct spatial derivatives $\p_q$ of $\Ps$ (and $\Qs$)
are not expected to be represented in this truncated
correspondence. They are thought to be represented by higher level
fields $P^{(\ell>1)}$ \cite{DaHeNi02} which we 
ignored in the $\s$-model and therefore this
disagreement is not surprising.
In the tensor equation (\ref{Tensor}), we have a term in the second
line which seems not to
be directly connected to a spatial derivative. However, if we assume
that the $\soso$ symmetry can be gauged and if we introduce an
$\soso$ valued zehnbein ${e_K}^p:=({e_k}^p,{e_{\bk}}^p)$ analogously
to the $SU(8)$ valued elfbein in \cite{deWiNi86}, we can write the
second line of the tensor equation (\ref{Tensor}) as
follows\footnote{$\delta^{KL}$
  denotes the second invariant tensor of $\soso$ which is similiarly
  defined as the first one, $\eta^{KL}$, but with
  $\delta^{\bk\bl}=+\delta^{kl}$.}
\be
     \delta_i^I\delta_{\bj}^J\left(\ns \p_p
     \left[{e_K}^{p}\delta^{KL}\Ps_{ILJ}\right]
   +e^{2\phis}{\Ps_I}^{KL}\Ps_{JKL}\right),
\ee
where indices have been raised with $\eta^{KL}$. Furthermore, the two
vector equations combine to a single one
\be
    0&=&-\ns\partial_{m}\left[{e_{K}}^m\eta^{KL}\ns^{-1}\Ps_{IL}\right]
   +e^{2\phis}{\Ps_I}^{KL} \ns^{-1}\Ps_{KL}\nn\\
   &&\quad +\ns e^{-\phis}\partial_{I}\left(\ns^{-1}e^{\phis}
   \partial_t\phis\right),
\ee
if we set $P_{ij}=P_{\bi\bj}=0$. This looks like an $\soso$-covariant
derivative $D_m$ with respect to {\em space} whereas before we only
considered $D_t$. This would imply that the term\footnote{The
   following equality shows that this term is precisely the 
   term which is also problematic in the Einstein equation of the
   maximal theory \cite{DaNi04}.}
\begin{eqnarray*}
    e^{2\phis}{\Ps_i}^{KL}\Ps_{\bj KL}
    &=e^{2\phi}\left(\Ps_{ikl}\Ps_{\bar{\jmath}kl} +
    \Ps_{i\bar{k}\bar{l}}\Ps_{\bar{\jmath}\bar{k}\bar{l}}
    -2\Ps_{\bar{k}il}\Ps_{l\bar{k}\bar{\jmath}} \right)
    =&
    -\frac{1}{2}\ns^{2}e^{-2\phi}\Omega_{(i|kl}\Omega_{\bar{\jmath})lk}
\end{eqnarray*}
should not be separated from the discussion of spatial
derivatives. The third line 
in the tensor equation (\ref{Tensor}) is an explicit spatial
derivative, which concludes the discussion of this equation.

Turning to the scalar equations we see that there are two different
mismatches. The first one is the common term in the second lines of
(\ref{Scalar1}) and (\ref{Scalar2}) whose value depends on the choice
of coordinate system and local Lorentz gauge. These freedoms could be used,
e.g., to let this term vanish or to fix
$v_a(t,\bx)=0$, as was suggested in \cite{KlNi04a}.\footnote{The
  quantity $v_a$ exactly is the spatial trace of the $D=11$ spin
  connection used in~\cite{KlNi04a}.}

The second mismatch in the scalar equations concerns the final line in
(\ref{Scalar1}). However, in order to account for this term we also
have the densitised lapse $\ns$ at our disposal which we can choose as
convenient as long as it does not vanish. Evidently, choosing
$\ns$ suitably we can cause this line to vanish identically at the fixed
spatial point $\bx$.

A fascinating possibility for taking the terms containing
spatial gradients into account in the correspondence 
was proposed in \cite{DaHeNi02}, where
it was suggested that they are related to some higher level fields of
the $\s$-model which have been truncated in our analysis. The
proper interpretation of these terms is still an open problem.

\end{subsection}

\end{section}

\begin{section}{Fermions and supersymmetry}
\label{ferm}

In this section, we extend our analysis to take into account the
fermionic degrees of freedom. We will in particular check that the
supersymmetry
transformations (\ref{SUSYvar}) can be stated in an $\soso$ covariant
form, which is necessary for the $K(DE_{10})$ covariance that is
conjectured to hold if all levels $\ell$ are fixed appropriately. We
start with the discussion of the fermionic variations, before we move
on to the bosonic fields. 

\begin{subsection}{$K(DE_{10})$ covariance of the fermionic
    transformations}
\label{fermtrm}

The supersymmetry transformation of the fermions $\lambda$ and
$\psi_M$ have been stated in equations (\ref{LSUSY}) and
(\ref{PsiSUSY}). In order to uncover the $\soso$ covariance, we have
to reparametrise the fermions as we have done with the bosons in
(\ref{DefBos1}) and (\ref{DefBos2}), where we explicitly break the
$SO(1,9)$ covariance again by treating the time component in a special
way
\be\label{chi}
\eps &:=&(\det(\hat{E}_m{}^a))^{-\frac{1}{2}}\epsilon,\nn\\
\chi_t &:=& (\det(\hat{E}_m{}^a))^{-\frac{1}{2}}\left(\psi_t -
     N\Gamma_0\Gamma^a\hat{\psi}_a\right),\nn\\
\chi_{\bi} &:=&
  (\det(\hat{E}_m{}^a))^{\frac{1}{2}}\left(\frac{1}{4}\Gamma_{\bi}\lambda - 
     \hat{\psi}_{\bi}\right),\nn\\
\chi &:=& (\det(\hat{E}_m{}^a))^{\frac{1}{2}}\left(\Gamma^a\hat{\psi}_a -
    \frac{1}{4}\lambda\right).
\ee
The hats denote, as in section \ref{sugra}, the projection onto the
orthonormal frame  $\hat{\psi}_a\equiv
\left.\hat{E}_a\right.^m\psi_m$.  The fermions in (\ref{chi}) can be assigned
$\soso$ transformation properties as follows: All spinors transform as
$16$-component Majorana spinors of the first $SO(9)$ factor and
trivially under the second $SO(9)$ except for $\chi_{\bi}$ which
transforms as a vector. This is consistent with the different
$SO(1,9)$ chiralities of the type~I fermions since single
 $(32\times 32)$ $\Gamma$-matrices,  defined in appendix
\ref{appendix}, intertwine between these two chiralities. The
$\soso$ representations considered here are the chiral half of the
representations of \cite{KlNi04a,KlNi06}.

Using the redefinitions (\ref{chi}),
(\ref{DefBos1}) and (\ref{DefBos2}) in the supersymmetry variations
(\ref{LSUSY}) and (\ref{PsiSUSY}), we arrive at the following
results in leading fermion order
\begin{subequations}\label{fermred}
\be
\delta_{\eps} \chi_t
        &=&
         \partial_t\eps
    +\frac{1}{4}\Gamma^{ij}\eps \Qs_{ij} +\frac{1}{3!}
    e^{\phis}\Gamma^{ijk}\Gamma^0\eps \Ps_{ijk}\nn\\
        &&+\ns e^{-\phis}\Gamma^{a}\Gamma^0\left\{-
        \partial_{a}\eps +\frac{1}{2}\eps\partial_{a}\phi
     +\frac{1}{2} \eps \vs_{a}\right\}\nn\\
     &&+\frac{1}{2}\ns
        e^{-\phis}\Gamma^{a}\Gamma^0\eps\left[\ns^{-1}\partial_{a}\ns
        -\partial_{a}\phi\right],\label{deltachit}\\
\delta_{\eps}\chi_{\bi}
        &=&
    -\frac{1}{2}\ns^{-1}\Gamma^{j}\Gamma^0\eps \Ps_{j\bi}
        +\frac{1}{2}\ns^{-1}e^{\phis}\Gamma^{jk} \eps
        \Ps_{\bi jk}\nn\\
     &&+e^{-\phis}\left\{-\partial_{\bi}\eps +\frac{1}{2}\eps
     \partial_{\bi}\phis\right\},\label{deltachii}\\
\delta_{\eps}\chi
    &=&-\frac{1}{2}\ns^{-1}\Gamma^0\eps\partial_t \phis
    -\frac{1}{3!}\ns^{-1}e^{\phis}\Gamma^{ijk}\eps \Ps_{ijk}\nn\\
    &&-e^{-\phis}\Gamma^{a}\Gamma^0\left\{-\partial_{a}\eps
    +\frac{1}{2}\eps \partial_{a}\phis +\frac{1}{2}\eps
    \vs_{a}\right\},\label{deltachi}
\ee
\end{subequations}
where we used the abbreviations defined in (\ref{dict}) and
(\ref{dict2}). We observe again that the $\soso$ structure is
preserved. From (\ref{deltachit}) one can also read off the beginning
of an extension of the $\soso$ covariance to $\KDE$ covariance along
the lines of \cite{KlNi04a,dBHP05a,DaKlNi06a,dBHP05b,DaKlNi06b} as we
now discuss.

The key to unravelling the $\KDE$ structure is to assume that the
bosonic $\s$-model can be extended to a
$\KDE$ gauge invariant and locally {\em supersymmetric} $D=1$ coset
model. This requires the introduction of
fermionic fields transforming in $\KDE$ representations. In such a
model there will be a superpartner $\chi_t$ to the lapse $n$ which
acts as the one-dimensional gravitino and therefore should transform
into a $\KDE$ covariant derivative of the supersymmetry parameter
\be\label{covfer}
\d_\eps \chi_t &=& \Abl \eps = \left(\p_t -\frac12 Q_{ij}J^{ij}-\frac12
Q_{\bi\bj}J^{\bi\bj}\right.\\
&&\quad\left. -\frac1{3!} P_{ijk}J^{ijk} -\frac1{2!} P_{\bi
  jk}J^{\bi jk}  -\frac1{2!} P_{\bi\bj k}J^{\bi\bj k}
-\frac1{3!}P_{\bi\bj\bk}J^{\bi\bj\bk} +\ldots\right)\eps\nn
\ee
in Borel gauge (\ref{pq}). By comparing this relation to
(\ref{deltachit}), we can read off the form the $\KDE$ generators take as a matrix representation on $\eps$. On the first two `levels', the result is
\begin{align}\label{drep}
J^{ij}\eps =& -\frac12\G^{ij}\eps,&\quad J^{\bi\bj}\eps =&\,0,&\nn\\
J^{ijk}\eps =& -\G^{ijk}\G^0\eps,&\quad J^{\bi jk}\eps =&\,0,&\quad J^{\bi\bj k}\eps =
0,\quad J^{\bi\bj\bk}\eps =0.
\end{align}
This implies that only two generators are represented non-trivially. In
\cite{DaKlNi06a,DaKlNi06b} it was demonstrated in the maximally
supersymmetric case that such restricted transformation
rules can be sufficient to prove that $\chi_t$ is a consistent {\em
  unfaithful} representation of $\KDE$. It follows from (\ref{covfer}) that $\eps$ has to transform in the same $\KDE$ representation. We now give the criterion for
establishing such a consistent representation and show that it is
satisfied in the present situation.

As shown in \cite{DaKlNi06b,Be89} the generators of the compact
subgroup of a Kac--Moody group can be written in terms of simple
generators $x_i = e_i-f_i$ (deduced from the Chevalley generators of
section~\ref{BWGL2}) and defining relations induced by the relations
satisfied by the generators $e_i$ and $f_i$. Working in an $\soso$
covariant formalism as we are doing here, the only consistency
relation to check turns out to be
\be\label{consrel}
\big[x_{10},\left[x_{10},x_3\right]\big] + x_3 =0,
\ee
where $x_{3}$ and $x_{10}$ are given in terms of the antisymmetric
generators (\ref{J0}) and (\ref{JIJK}) via \cite{KlNi04a}
\be
x_3 &=& J^{3\,4} + J^{\bar{3}\,\bar{4}},\\
x_{10} &=& \frac12\left( J^{1\,2\,3} + J^{\bar{1}\,2\,3} +
J^{1\,\bar{2}\,3} + J^{1\,2\,\bar{3}} + J^{\bar{1}\,\bar{2}\,3} +
J^{\bar{1}\,2\,\bar{3}} + J^{1\,\bar{2}\,\bar{3}} +
J^{\bar{1}\,\bar{2}\,\bar{3}}\right).\nn
\ee
By substituting these generators into the consistency relation
(\ref{consrel}) in a specific matrix representation acting on a
vector space $V$, one can check whether $V$ is a
representation space of $\KDE$.\footnote{We reiterate that we assume
  $\soso$ covariance of all generators --- if this is not
  guaranteed there are additional relations which need to be
  verified.}

The specific expressions for $J^{ij}$, $J^{\bi\bj}$ and $J^{IJK}$
found in (\ref{drep}) satisfy the relation (\ref{consrel}) as can be
checked by straight-forward $\G$-algebra. In terms of $(32\times 32)$
matrices the representation matrices $\G^{ij}$ and $\G^{ijk}\G^0$ are
block-diagonal and so act consistently on the projected
$16$-dimensional (chiral) spinor $\chi_t$. Naturally, there is also a
representation on the other $16$ components whose representation
matrices are given by
\begin{align}\label{drep2}
J^{ij}\eta =&\, 0,&\quad J^{\bi\bj}\eta =& -\frac12\G^{ij}\eta,&\nn\\
J^{ijk}\eta =&\, 0,&\quad J^{\bi jk}\eta =&\, 0,&\quad J^{\bi\bj k}\eta =
0,\quad J^{\bi\bj\bk}\eta = -\G^{ijk}\G^0\eta.
\end{align}
Therefore, there are two inequivalent $16$-dimensional unfaithful
spinor representations of $\KDE$ as already anticipated in
\cite{KlNi06}.\footnote{The $32$-dimensional unfaithful spinor representation of $K(E_{10})$ \cite{dBHP05a,DaKlNi06a} decomposes into the sum of these two inequivalent $16$-dimensional spinor representations of the $\KDE$ subgroup of $K(E_{10})$.} One can write down a similar consistent unfaithful
representation of $\KDE$ on $\chi_{\bi}$ which has dimension $144$; to
deduce its transformation laws one needs to rewrite the fermionic
equation of motion and interpret this as a $\KDE$ covariant derivative
of $\chi_{\bi}$ \cite{DaKlNi06a,DaKlNi06b}.

Having discussed the first few terms in (\ref{deltachit}), we now briefly
explore the remaining structure of
eqs.~(\ref{fermred}). If
we compare the two scalar equations of motion (\ref{Scalar1}) and
(\ref{Scalar2}) with the corresponding supersymmetry transformations
(\ref{deltachit}) and (\ref{deltachi}), the similarities are striking:
Apart from the global factor of $-\ns$, the second lines of
(\ref{deltachit}) and (\ref{deltachi}) completely agree as in
(\ref{Scalar1}) and (\ref{Scalar2}) and the third line of
(\ref{deltachit}) has the same structure as the one in
(\ref{Scalar1}). The final interpretation of these lines is still an
open problem. However, the proposals which we discussed in section
\ref{Comp} can be equally applied here. 

\end{subsection}

\begin{subsection}{Supersymmetry transformation of the
  bosons}\label{boson}

The redefined variables $\Ps$ and $\Qs$ defined in
(\ref{dict}) and (\ref{dict2}) have been assigned $\soso$
transformation properties. Given the $\soso\subset\KDE$ fermions of
the preceding section it is natural to study the $\soso$ and $\KDE$
transformation properties of $\Ps$ and $\Qs$ after a supersymmetry
transformation $\delta_{\eps}$.

We recall the definitions of $\Ps_{i\bj}$, $\Qs_{ij}$ and
$\Qs_{\bi\bj}$ from (\ref{dict}), where the latter two play the role
of $\soso$ gauge connections in the equations of motion (\ref{Tensor})
and (\ref{P2}). It will prove useful to define analogous quantities by
replacing the time derivative $\p_t$ by the supersymmetry variation
$\delta_\e$ in (\ref{dict}) and hence get\footnote{As in section
  \ref{redefs}, there is no distinction between barred and unbarred
  frame indices at this point.}
\begin{subequations}\label{LaSiSib}
\begin{eqnarray}
    \Lambda_{i\bar{\jmath}}&:=&{e_{(i}}^m\delta_\epsilon
    e_{m|{\bar{\jmath}})}
    -\frac{1}{2}{e_{i}}^m{e_{\bar{\jmath}}}^n\delta_\epsilon
    B_{mn}\label{La},\\
    \Sigma_{ij}&:=&{e_{[i}}^m\delta_\epsilon e_{m|j]}
    +\frac{1}{2}{e_{i}}^m{e_{j}}^n\delta_\epsilon
    B_{mn}\label{Si},\\
    \Sigma_{\bar{\imath}\bar{\jmath}}
   &:=&{e_{[{\bar{\imath}}}}^m\delta_\epsilon
    e_{m|{\bar{\jmath}}]}
    -\frac{1}{2}{e_{\bi}}^m{e_{\bj}}^n\delta_\epsilon
    B_{mn}.\label{Sib}
\end{eqnarray}
\end{subequations}
Substituting in the supersymmetry variations (\ref{SUSYvar}) as well as the redefinitions of the bosons (\ref{DefBos1}) and the fermions (\ref{fermred}), we find explicitly
\be\label{explizit}
	\Lambda_{i\bar{\jmath}} = -i\eps\Gamma_i\chi_{\bj}.
\ee
Furthermore, in analogy to the $\soso$ covariant derivative $D_t$
defined in (\ref{sosocov}), we define an $\soso$ covariant
supersymmetry transformation $\Sdelta_\eps$ by adding a local (in
time) and field dependent $\soso$ gauge transformation
$\delta_\Sigma$ to the supersymmetry variation $\delta_\eps$
\be\label{Sdelta}
    \Sdelta_\eps = \delta_\eps +\delta_\Sigma,
\ee
as was done in the $SU(8)$ case in \cite{deWiNi86}.\footnote{As $\Sigma=(\Sigma_{ij},\Sigma_{\bi\bj})$ is of order two in fermions and as we have neglected higher order fermion contributions throughout the paper, the introduction of the covariant supersymmetry transformation $\Sdelta_\eps$ does not affect the discussion of the fermions in section \ref{fermtrm}.} With the
definitions (\ref{omega}) and (\ref{dict}), a short calculation yields
the identities
\begin{subequations}\label{deltaP0}
\be
    \Sdelta_\eps \Ps_{i\bar{\jmath}}= \delta_\eps
    \Ps_{i\bar{\jmath}} +\Sigma_{ik}P_{k\bj} +
    \Sigma_{\bj\bk}P_{i\bk}
    &=&D_t \Lambda_{i\bar{\jmath}}
    \label{deltaP},\\
\Sdelta_\eps \Qs_{ij} = \delta_\eps \Qs_{ij} +2\Sigma_{[i|l}\Qs_{l|j]}
-\partial_t \Sigma_{ij}
&=&  2\Ps_{[i|\bar{l}}\Lambda_{j]\bar{l}}\label{deltaQ},\\
\Sdelta_\eps \Qs_{\bar{\imath}\bar{\jmath}} = \delta_\eps
    \Qs_{\bar{\imath}\bar{\jmath}}
    +2\Sigma_{[\bar{\imath}|\bar{l}}\Qs_{\bar{l}|\bar{\jmath}]}
-\partial_t \Sigma_{\bar{\imath}\bar{\jmath}}
&=& 2\Ps_{l[\bar{\imath}}\Lambda_{l|\bar{\jmath}]},
\label{deltaQquer}
\ee
\end{subequations}
where the gauge fields $\Qs_{ij}$ and $\Qs_{\bi\bj}$ transform with
explicit time derivatives of the gauge transformation parameters
$\Sigma_{ij}$ and $\Sigma_{\bi\bj}$ as usual.

For the fields defined in (\ref{dict2}), we get
\begin{subequations}\label{deltaP2}
\be
    \ns e^{-2\phi} \Sdelta_\eps
     \left(\ns^{-1}e^{2\phi}\Ps_{ijk}\right)
     &=& -3\Lambda_{[i|\bar{l}}   \Ps_{\bar{l}|jk]}
     -\frac{3}{2}\partial_{[i}\Sigma_{jk]},
     \label{deltaPijk}\\
    \ns e^{-2\phi} \Sdelta_\eps \left(\ns^{-1}e^{2\phi}\Ps_{\bi jk}\right)
      &=&
    -\Lambda_{l\bi}\Ps_{ljk}
      -2\Lambda_{[j|\bar{l}}\Ps_{k]\bi\bar{l}} \nn\\
    &&-\frac{1}{2}\partial_{\bi}\Sigma_{jk} +\partial_{[j}\Lambda_{k]\bi},
      \label{deltaPaquerjk}\\
    \ns e^{-2\phi} \Sdelta_\eps
    \left(\ns^{-1}e^{2\phi}\Ps_{i\bar{\jmath}\bar{k}}\right)
    &=&
    -\Lambda_{i\bar{l}}\Ps_{\bar{l}\bar{\jmath}\bar{k}}
    -2\Lambda_{l[\bar{\jmath}}\Ps_{\bar{k}]il}\nn\\
    &&+\frac{1}{2}\partial_{i}\Sigma_{\bar{\jmath}\bar{k}}
    -\partial_{[\bj}\Lambda_{|a|\bar{k}]},
\label{deltaPajquerkquer}\\
    \ns e^{-2\phi} \Sdelta_\eps
     \left(\ns^{-1}e^{2\phi}\Ps_{\bar{\imath}\bar{\jmath}\bar{k}}\right)
     &=&  -3\Lambda_{l[\bar{\imath}}
     \Ps_{|l|\bar{\jmath}\bar{k}]}
     +\frac{3}{2}\partial_{[\bar{\imath}}\Sigma_{\bar{\jmath}\bar{k}]}.
     \label{deltaPiquerjquerkquer}
\ee
\end{subequations}
Again, we observe the $SO(9)\times SO(9)$ covariance. The
appearance of spatial derivatives of the $\soso$ transformation parameter
$\Sigma$ indicates that we should gauge the symmetry group $\soso$
with respect to space-time in fact by introducing also an $\soso$
derivative $D_m$ as discussed in~\ref{Comp}. However, this is not the
way we want to pursue; instead we now study supersymmetry transformations in the
$\DEC$ coset structure. \\

In the derivation of the coset equations of motion in section
\ref{EOMcoset}, we started with an element $\cV$ in the group coset
$DE_{10}/K(DE_{10})$ and considered variations of $\cV$ under a
general variation $\d$.
As the supersymmetry operator is also realized as a derivative
operator $\delta_\eps$, which commutes with the time derivative
$\partial_t$, we can use the same chain of arguments of
section \ref{EOMcoset} to derive the supersymmetry variation of $\cP$
and $\cQ$. This means we first decompose the
$\text{Lie}(DE_{10})$ valued expression
\be
\delta_\eps \cV \cV^{-1} = \clAmbda + \csIgma
\ee
into generators $\csIgma\in\text{Lie}(K(DE_{10}))$ and
$\clAmbda\in\text{Lie}(DE_{10}/K(DE_{10}))$. Then, we parametrise
$\csIgma$ and $\clAmbda$ similiarly to (\ref{pq}) in a level
decomposition truncated for $\ell\geq 2$, {\em i.e.}
\be\label{ls}
\clAmbda &=& \delta_\eps \vp T + \lAmbda_{i\bj}S^{i\bj}
     +\frac{1}{3!}e^{\vp} \lAmbda_{IJK}S^{IJK},\nn\\
\csIgma &=& \frac{1}{2}\sIgma_{ij}J^{ij}  +\frac{1}{2}\sIgma_{\bi\bj}J^{\bi\bj}
   +\frac{1}{3!}e^{\vp} \lAmbda_{IJK}J^{IJK},
\ee
where we again work in Borel gauge.
Finally, from the fact that both derivative operators commute 
it follows that we get the variations
after projecting on the $\ell=0$ generators $S^{i\bj}$, $T$  and the gauge
orbit generators $J^{ij}$ and $J^{\bi\bj}$
\begin{subequations}\label{CosdeltaP0}
\be
\Sdelta_\eps (\p_t\vp) &=& \d_\eps (\p_t\vp) = \p_t(\d_\eps\vp),\\
        \Sdelta_\eps P_{i\bar{\jmath}} &=&\delta_\eps
        P_{i\bar{\jmath}} +\sIgma_{ik}P_{k\bar{\jmath}}
        +\sIgma_{\bar{\jmath}\bar{k}}P_{i\bar{k}}
        = D_t\lAmbda_{i\bar{\jmath}},
     \\
     \Sdelta_\eps Q_{ij} &=& \delta Q_{ij} +2\sIgma_{[i|k}Q_{k|j]}
        -\partial_t\sIgma_{ij}
      = 2P_{[i|\bar{l}}\lAmbda_{j]\bar{l}}, \\
    \Sdelta_\eps Q_{\bar{\imath}\bar{\jmath}} &=& \delta
        Q_{\bar{\imath}\bar{\jmath}}
        +2\sIgma_{[\bar{\imath}|\bar{k}}Q_{\bar{k}|\bar{\jmath}]}
        -\partial_t\sIgma_{\bar{\imath}\bar{\jmath}}
    = 2P_{k[\bar{\imath}}\lAmbda_{k|\bar{\jmath}]},
\ee
\end{subequations}
where we have used the covariant derivative $D_t$ as in
(\ref{CosTensor}). With the covariant supersymmetry transformation
$\Sdelta_\eps$ we can write the equations resulting from the
projection onto the $\ell= 1$ generators $S^{ijk}$, $S^{\bi jk}$,
$S^{i\bj \bk}$ and $S^{\bi\bj\bk}$ as
\begin{subequations}\label{CosdeltaP2}
\be
    \Sdelta_\eps P_{ijk}    - 3 \lAmbda_{[i|\bar{l}} P_{\bar{l}|jk]}
    &=&D_t\lAmbda_{ijk} - 3 P_{[i|\bar{l}} \lAmbda_{\bar{l}|jk]},  \\
\Sdelta_\eps P_{\bi jk}-\lAmbda_{l\bi} P_{ljk} +2 \lAmbda_{[j|\bar{l}}
  P_{k]\bi\bar{l}}
&=& D_t \lAmbda_{\bi jk}-P_{l\bi} \lAmbda_{ljk} +2 P_{[j|\bar{l}}
  \lAmbda_{k]\bi\bar{l}},\\
\Sdelta_\eps P_{i\bar{\jmath}\bar{k}} -2\lAmbda_{m[\bar{\jmath}}
  P_{\bar{k}]im} + \lAmbda_{i\bar{l}} P_{\bar{l}\bar{\jmath}\bar{k}}
&=& D_t \lAmbda_{i\bar{\jmath}\bar{k}} -2P_{m[\bar{\jmath}}
  \lAmbda_{\bar{k}]im} + P_{i\bar{l}}
\lAmbda_{\bar{l}\bar{\jmath}\bar{k}},\quad\quad
\\
\Sdelta_\eps P_{\bar{\imath}\bar{\jmath}\bar{k}} -3
\lAmbda_{l[\bar{\imath}} P_{|l|\bar{\jmath}\bar{k}]}
&=&D_t \lAmbda_{\bar{\imath}\bar{\jmath}\bar{k}} -3 P_{l[\bar{\imath}}
  \lAmbda_{|l|\bar{\jmath}\bar{k}]}.
\ee
\end{subequations}
It should be noted that an inclusion of higher level terms $\ell\geq
2$ in the expansions (\ref{pq}) and (\ref{ls}) above does not alter
the equations (\ref{CosdeltaP0}) and (\ref{CosdeltaP2}) in
contradistinction to the equations of motion (\ref{eoml0}) and
(\ref{eoml1}). This is a general property of the Borel gauge
\cite{DaKlNi06b}.

As we have identified the supergravity variables $\Ps$ and $\Qs$ at a
fixed spatial point $\bx$ with the coset variables $P$ and $Q$ in section
\ref{Comp}, a comparison of the equations (\ref{deltaP0}) and
(\ref{CosdeltaP0}) forces us to identify $\Lambda_{i\bj}$ with
$\lAmbda_{i\bj} $ and
$(\Sigma_{ij},\Sigma_{\bi\bj})$ with $(\sIgma_{ij},\sIgma_{\bi\bj})$,
respectively. Using the explicit form of 
$\lAmbda_{i\bj}$ in terms of fermion bilinears (\ref{explizit}) and the $\KDE$
transformation rules for the fermions deduced in
section~\ref{fermtrm}, we could in principle compute
$\lAmbda_{IJK}$ from this by comparing it with a $\KDE$
transformation of the coset representation. We can obtain an
independent answer for $\lAmbda_{IJK}$ by comparing with
supergravity. It is not guaranteed that the two answers will agree. In
\cite{DaKlNi06b} a similar analysis was carried out in the maximal
$D=11$ supergravity context and some but not all expressions for the
analogues of $\lAmbda_{IJK}$ agree. We 
take this as an indication that there is a disparity between the
unfaithful, finite-dimensional fermionic $\KDE$ representation and the
infinite-dimensional coset representation, which is in conflict with
supersymmetry on the coset side.

\end{subsection}

\end{section}

\begin{section}{Discussion}
\label{concl}

In this paper we have rewritten the bosonic and fermionic fields
of pure type~I supergravity in terms of variables which we
assigned to representations of $\soso$ (cf. (\ref{dict}),
(\ref{dict2}) and (\ref{chi})). The relevant
bosonic representations are identical to those that arise in the
$D_9$ level decomposition of $\DE$ on the levels $\ell=0,1$
as shown in section~\ref{BWGL2}. In section~\ref{fermtrm} we also
showed that some of the relevant fermionic representations of $\soso$
can be consistently extended to unfaithful representations of
$\KDE$. 

At the dynamical level we demonstrated that the bosonic
equations of pure type~I supergravity
in this parametrisation (evaluated at a fixed spatial point)
coincide with those derived from a simple $D=1$ non-linear
$\s$-model on $\DEC$ truncated consistently beyond $\ell=1$ up to
a number of terms which can either be gauged away or can be argued
to be of the form of (generalised) spatial gradients, see
section~\ref{Comp}.\footnote{Our analysis was always at the level
  of the equations of motion and not at the level of the action.}
It would be very interesting to see whether
the full $D=10$ equations can be cast in $\soso$ covariant form,
analogous to the treatment in \cite{deWiNi86}. Our focus was not
on this question but rather if we can extend the (partial) $\soso$
covariance to a (partial) $\KDE$ covariance to further test the
ideas of \cite{DaHeNi02}. As pointed out for example below
(\ref{eoml1}) the truncated coset model equations of motion
include terms which are part of a $\KDE$ covariant formulation and
these terms agree with identical terms in the supergravity
equations (\ref{P2}). A similar phenomenon was observed for the
supersymmetry variation of certain fermionic fields, see
(\ref{covfer}). 

We consider our results as evidence that
$\KDE$ might be a dynamical symmetry of the pure type~I theory.
There are also a number of conundrums related to our analysis,
some of which were already hinted at.

As is well known, the pure type~I supergravity theory is not anomaly
free. However, by adding appropriate vector multiplets
\cite{ChMa82} the anomalies can be cancelled
\cite{GrSchw84b,GrSchw85a}. The possibilities which are realised
as string theory low energy effective theories are those which
have vector multiplets transforming as Yang--Mills fields of
either $SO(32)$ or $E_8\times E_8$. The inclusion of these
non-Abelian symmetries in the context of Kac--Moody symmetries is
poorly understood. Augmenting the theory (\ref{AI}) by
Abelian or non-Abelian vector fields
changes the associated cosmological billiard from $\DE$ to a group
called $BE_{10}$ \cite{DaHe01} but, at the level of coset model,
$BE_{10}$ is not appropriate for accommodating more than a single
Abelian vector field. In \cite{SchnWe04} multiple Abelian vector
fields were added to the pure type~I theory by increasing the
rank of the Kac--Moody symmetry and changing the real
form.\footnote{See also
  \cite{BrGaGaHe05} for a discussion of a non-maximal theory with
  vector fields obtained by orbifolding the maximal theory. In
  this analysis $\DE$ and further extensions of it similar to those of
  \cite{SchnWe04} appear.} A proper understanding of the
non-Abelian symmetries from a Kac--Moody algebraic point of view
is lacking at the moment.

Another interesting challenge is to extend the bosonic $\DEC$
$\s$-model of (\ref{l}) to a locally supersymmetric model in $D=1$. With the
fermionic representations employed in  this paper, it appears
impossible to construct such a model. In fact, in the present situation
there is no non-vanishing combination that can be
constructed from terms bilinear in the fermions (\ref{chi})
of the form $\lAmbda_{\bi\bj k}$ whereas such an expression 
necessarily appears in the supersymmetry variation of the $\ell=1$
field $P_{\bi\bj k}$ due to (\ref{deltaP2}) and (\ref{CosdeltaP2}).

Finally, it is crucial to bring the `gradient conjecture' of
\cite{DaHeNi02} back into view. According to this conjecture the
$\s$-model can capture the {\em full} dynamics in a neighbourhood
of the fixed spatial point $\bx$ by translating the information
about all spatial gradients of the supergravity fields into higher
level degrees of freedom of the $\s$-model. As discussed in
section~\ref{Comp}, the concrete realisation of this translation is still
an open problem.

\mbox{}\\

{\bf Acknowledgements}\\
We are grateful to H.~Nicolai for discussions on the results presented
here. CH would like to thank the Studienstiftung des deutschen Volkes for
financial support. This research was partly supported by the European
Research and Training Networks `Superstrings' (contract number
MRTN-CT-2004-512194).

\end{section}

\renewcommand{\thesection}{\Alph{section}}
\setcounter{section}{0}

\begin{section}{Conventions for $\G$-matrices}
\label{appendix}

We use the same conventions for $\G$-matrices as
\cite{KlNi04a} which we summarise for completeness.

The $(32\times 32)$ real $\G$-matrices $\G^A$ for $A=0,\ldots,10$ of
$SO(1,10)$ are defined in terms of the real symmetric $(16\times 16)$
$\g$-matrices $\g^i$ ($i=1,\ldots,9$) of $SO(9)$ by
\be
\G^0=\left(\begin{array}{cc}0&-{\bf 1}_{16}\\{\bf 1}_{16}&0\end{array}\right)
\; , \;
\G^{10}=\left(\begin{array}{cc}{\bf 1}_{16}& 0\\0& -{\bf
    1}_{16}\end{array}\right)
\; , \;
\G^i=\left(\begin{array}{cc}0&\g^i\\ \g^i&0\end{array}\right).
\ee
The matrix $\G^0$ is the charge conjugation matrix in $D=11$.
After descending to $SO(1,9)$ the matrix $\frac12({\bf
  1}_{32}\pm\G^{10})$ serves as the projector on the two chiral
spinors in $D=10$. The type~I fermions $\psi_M$ and $\lambda$ discussed in
the paper have been projected from $32$-component spinors to
opposite chiralities.

\end{section}

\end{document}